**Controllable growth of centimeter-size 2D perovskite heterostructural single crystals for highly narrow dual-band photodetectors**

Jun Wang[1#], Junze Li[1#], Shangui Lan[1], Chen Fang[1], Hongzhi Shen[1] and Dehui Li[1, 2*]

[1]*School of Optical and Electronic Information, Huazhong University of Science and Technology, Wuhan, 430074, China;*

[2]*Wuhan National Laboratory for Optoelectronics, Huazhong University of Science and*

*Technology, Wuhan, 430074, China;*

[#] Those authors contribute to this work equally.

*Correspondence to: Email: dehuili@hust.edu.cn.

## Abstract

Two-dimensional (2D) organic-inorganic perovskites have recently attracted increasing attention due to their great environmental stability, remarkable quantum confinement effect and layered characteristic. Heterostructures consisting of 2D layered perovskites are expected to exhibit new physical phenomena inaccessible to the single 2D perovskites and can greatly extend their functionalities for novel electronic and optoelectronic applications. Herein, we develop a novel solution method to synthesize $(C_4H_9NH_3)_2PbI_4/(C_4H_9NH_3)(CH_3NH_3)Pb_2I_7$ single-crystals with the centimeter size, high phase purity, controllable junction depth, high crystalline quality and great stability for highly narrow dual-band photodetectors. On the basis of the different lattice constant, solubility and growth rate between $(C_4H_9NH_3)_2PbI_4$ and $(C_4H_9NH_3)(CH_3NH_3)Pb_2I_7$, the newly designed synthesis method allows to first grow the $(C_4H_9NH_3)_2PbI_4$ guided by the self-assembled layer of the organic cations at the water-air interface and subsequently the $(C_4H_9NH_3)(CH_3NH_3)Pb_2I_7$ layer is formed via diffusion process. Such growth process provides an efficient away for us to readily obtain the $(C_4H_9NH_3)_2PbI_4/(C_4H_9NH_3)(CH_3NH_3)Pb_2I_7$ single-crystals with various thickness and junction depth by controlling the concentration, reaction temperature and time. The formation of heterostructures has been verified by X-ray diffraction, cross-section photoluminescence and reflection spectroscopy with the estimated junction width below 70 nm. Photodetectors based on such heterostructural single crystal plates exhibit extremely low dark current ($\sim 10^{-12}$ A), high on/off current ratio ($\sim 10^3$), and highly narrow dual-band spectral response with a full-width at half-maximum of 20 nm at 540 nm and 34 nm at 610 nm due to the high crystalline quality of the synthesized heterostructures and extremely large resistance in the out-of-plane direction leading to the efficient control of photogenerated carrier collection. In particular, the synthetic strategy is general for other 2D perovskites and the narrow dual-band spectral response with all full-width at half-maximum <40 nm can be continuously tuned from red to blue by properly changing the halide compositions. Our findings not only provide an efficient synthetic approach with great simplicity to create 2D perovskite based heterostructural single crystals for investigating the physical processes in those heterostructures, but also offer an alternative strategy to achieve optical-filterless narrow dual-band photodetectors in the entire visible range for multicolor optical sensing.

**Introduction**

Organohalide lead perovskites (OHIPs) have been extensively studied in the past few years partially due to the rapid surge of the power conversion efficiency (PCE) of the perovskite solar cells to a certified efficiency higher than 22 %[1-3]. In addition, benefiting from the large light absorption coefficient in the UV-vis range, the high carrier mobility and long carrier diffusion length, OHIPs-based photodetectors, lasers and light-emitting devices have been also demonstrated with excellent performance[4-7]. Nevertheless, OHIPs are extremely unstable against moisture, heat and light illumination, which severely limits their applications in practice[8,9].

To address the long-term stability issue of 3D perovskites mentioned, two-dimensional (2D) organic-inorganic perovskites have emerged as alternative materials with great environmental stability and less toxicity due to the reduced lead content[10-12]. 2D perovskites have a general chemical formula of $(A)_NB_{N-1}M_NX_{3N+1}$, in which A and B are organic cations, M is a divalent metal and X is a halide anion[13,14]. The integer N represents the number of $[MX_6]^{4-}$ octahedron layers sandwiched between two layers of A spacer cations. Cation A is usually hydrophobic alkyl chain, which can efficiently prevent the $[MX_6]^{4-}$ octahedrons from being directly contacted with moisture in ambient and thus improve the stability. Additionally, $[MX_6]^{4-}$ octahedron layers are separated by the organic chains which act as dielectric layers and thus quantum-well structures are naturally formed leading to the giant quantum confinement effect and large exciton binding energy[15,16]. Importantly, the weak Van der Waals interaction between inorganic layers and organic layers allows us to obtain ultrathin perovskite layers by mechanical exfoliation method and subsequently integrate with other 2D layered materials such as graphene and transition metal dichalcogenides to achieve the desired material properties as demanding[17,18]. While the bandgap of the 2D perovskites can be easily modified via adjusting the number N or chemical compositions, the dielectric environment and quantum confinement offer additional degrees of freedom to tune the optical and optoelectronic properties of the materials, all of which would provide tremendous flexibility for the potential electronic and optoelectronic applications such as photodetectors, lasers and light emitting devices[17-19]. Up to date, the solar cells based on 2D perovskites has been demonstrated with greatly improved stability and fairly good PCE of 12.9 %[20].

Heterostructures consisting of two or more materials with different energy landscape are expected to exhibit new physical phenomena inaccessible to the single material system and can greatly extend their functionalities for novel electronic and optoelectronic applications[21,22]. To fully explore their potential applications, rational design and controllable synthesis of heterostructures are indispensable. A variety of methods including molecular beam epitaxy and metal-organic vapor phase epitaxy have been used to fabricate large-scale III-V and II-IV group based heterostructures while alignment transfer technique and chemical vapor deposition have been adopted to prepare the heterostructures or multi-heterostructures based on graphene and other 2D layered materials[23,24]. The advancement of the preparation methods greatly prompts the studies of heterostructures in field-effect transistors, quantum-well lasers, room-temperature magnetoresistance and light emitting devices[25,26]. Owing to the layer characteristic in nature, 2D perovskites are inherently ideal candidates for heterostructures and 2D perovskites with different chemical compositions or different N number can seamlessly form heterostructures without well-width fluctuation, interface defects and needless to fulfill the strict requirement of the lattice match. Large scale 2D perovskite based lateral and vertical heterostructures have been prepared with great simplicity and flexibility by a combination of the solution and gas-solid phase intercalation method[27]. Nevertheless, the prepared heterostructures are not single crystals and the intercalation process would introduce grains leading to the poor performance of the photodetectors based on them. Therefore, controllable synthesis of large-scale single crystal heterostructures based on 2D perovskites is urgently required in order to improve the device performance, investigate the physical processes and explore their potential applications.

Wavelength-selective photodetectors with narrow spectral response have important applications in the field of biomedical sensing, imaging, defense and machine vision[28-30]. Up to date, the narrowband photodetectors have been achieved via (1) combining broadband photodetectors with bandpass filters, (2) intentionally design active absorption materials with narrowband[31], (3) adopting plasmonic effect to enhance absorption in a designed wavelength range[32] and (4) manipulating the photogenerated carrier collection efficiency[29]. Narrowband photodetectors based on the above strategies have been developed for color discrimination with the full width at half-maximum (FWHM) ranging from 20 nm to 100 nm[33,34]. In particular, wavelength-selective photodetectors with dual and multiple narrowband spectral response in the entire visible range would be essential in multicolor image technology, which have attracted great attention due to their potential applications in a variety of areas[35-37]. Nevertheless, investigations on narrow dual-band photodetectors are still in its fancy stage. Here, we report on a novel solution method to synthesize $(C_4H_9NH_3)_2PbI_4/(C_4H_9NH_3)_2(CH_3NH_3)_2Pb_2I_7$ single-crystals with the centimeter size, controllable junction depth, high crystalline quality and great stability for highly narrow dual-band photodetectors. By changing the chemical compositions, the narrow dual-band spectral response can be readily tuned in the entire visible range with FWHM smaller than 35 nm, extremely low dark current and high on-off ratio.

Figure 1a shows the schematic illustrations of a general solution method for scalable synthesis of $(BA)_2(MA)_{N-1}Pb_NI_{3N+1}$ (N=1-5) perovskite films and plates, where BA=n-$CH_3(CH_2)_3NH_3^+$ and MA=$CH_3NH_3^+$[10]. PbO powder was first dissolved in a mixture of HI and $H_3PO_2$ aqueous solution resulting in a bright yellow solution by heating to boiling under constant magnetic stirring. Subsequently, MAI powder was added to the resultant solution to form a MAPbI$_3$ solution and 2D perovskite $(BA)_2(MA)_{N-1}Pb_NI_{3N+1}$ with different N value could also be obtained by adding the BAI solution with carefully controlling the stoichiometry of MAI and BAI[38]. Due to the similar thermodynamics property for N>2 and the formation of 3D MAPbI$_3$ at the first stage before 2D perovskites, it is difficult or even impossible to controllably synthesize 2D perovskite based $(BA)_2PbI_4/(BA)_2(MA)Pb_2I_7$ heterostructures with high phase purity using this method. Rather, mixture of 2D perovskites with different N value is expected to be created.

To solve this issue, we hypothesized a new method to synthesize $(BA)_2PbI_4/(BA)_2(MA)Pb_2I_7$ single crystals based on the different thermodynamics properties, lattice constant and growth rate between $(BA)_2PbI_4$ (N=1) and $(BA)_2(MA)Pb_2I_7$ (N=2) and the different solubility of the 2D perovskite members with different N value, where the reaction sequence is reversed: BAI solution was first injected into the yellow solution at 80 °C and MAI powder was added subsequently (Fig.1b). As has been reported previously, the solubility of 2D perovskites series in the aqueous solution increases with the increase of N value and finally reaches a maximum solubility for 3D MAPbI$_3$ (N=∞)[39-41]. When the temperature is reduced to a certain value (~ 75 °C), the N=1 2D perovskites start to quickly precipitate from the solution due to the lower solubility in the aqueous solution. Subsequently, with further slightly decreasing the temperature, the N=2 2D perovskites could form onto the outside of the synthesized N=1 perovskite plates by inserting the MA cations into the center of $[PbI_6]^{4-}$ octahedrons of the N=1 perovskite plates and releasing BAI molecules away.

The schematic illustration of the detailed growth mechanics is displayed in Fig. 1c. The growth of N=1 perovskites takes place anisotropically with a much faster growth rate along the in-plane direction at the water-air interface where a self-assembled layer of the organic cations is formed and serves as the soft template guiding the formation of the perovskite underneath (Fig. 1c) [ref]. This is supported by the fact the as-growth plates always locate at the surface of the solution in our experiments (Supplementary Fig. 1). For such growth mechanism, the growth rate for N=1 is faster than that for N=2 along the in-plane direction due to the higher solubility of methylammonium compared with that of butylammonium [ref]. In addition, due to the presence of large lattice mismatch between N=1 and N=2 along the in-plane direction (Supplementary Fig. 2), it would be likely to prohibit the growth of N=2 in the lateral direction. Combining the faster growth rate for N=1 together with the large lattice mismatch between N=1 and N=2, we expect that the growth of N=1 layer occurs in ahead of the N=2 and no mixture of N=1 and N=2 in the in-plane direction is formed as shown in Fig. 1c. After the formation of the N=1, the N=2 layer is grown outside of the synthesized N=1 perovskite plates by diffusion controlled process at the slightly lower temperature. By this way, it is expected that the thickness of the outer N=2 layer is able to be well controlled as demanding by properly tuning the maintaining time at 75 °C and highly pure N=1/N=2 single crystal heterostructures can be obtained by carefully controlling the reaction conditions without introducing N>2 perovskite phases since N>2 phases have much higher solubility and thus lower growth rate compared with N=1 and N=2 phases [ref].

Figure 2a shows the picture of the as-synthesized centimeter-size N=1/N=2 heterostructural single crystal plates (maintaining the reaction at 75 ºC for 240 min with MAI concentration of 1Mmol and the BAI to MAI ratio of 2:3) with the red color corresponding to the color of N=2[27], which indicates the exterior of the heterostructural plates consists of N=2 layer. Nevertheless, when we mechanically exfoliated the plates by scotch tape, the interior of the plates exhibits yellow color corresponding to N=1[27], which reveals the formation of the N=1/N=2 (Fig. 2b). Scanning electron microscopy image shows the surface of the heterostructure plates synthesized by this method is rather smooth (inset of Fig. 2a). The X-ray diffraction pattern of the resultant plates shows that all diffraction peaks can be well indexed to (00k) peaks of N=1 and (0k0) peaks of N=2 without any other unassigned peaks, which suggests the phase purity and high crystalline

quality of the heterostructures (Fig. 2b)[42]. For comparison, the XRD diffraction patterns for pure

N=1 and N=2 2D perovskites[39] synthesized by solution method have been also included in Fig. 2c.

The room-temperature photoluminescence (PL) spectra (Fig. 2d) of the resultant plates show that the emission peaks locate at 520 nm (green band) and 590 nm (yellow band), consistent with the exciton emission peak position of the N=1 and N=2 2D perovskite films, which further verifies the formation of the N=1/N=2 heterostructures[27].

Spatially resolved PL mapping images suggest that the as-grown N=1/N=2 heterostroctural

single crystals have high spatial uniformity (Fig. 2e, f) while the spatially resolved cross-section

PL mapping at 520 nm and 580 nm indicates the existence of the distinct interface between N=1

and N=2 portion (Fig. 2g). The decrease of the emission intensity near the junction in Fig. 2g is

due to the type-II band alignment between N=1 and N=2, which leads to the separation of the

photogenerated carriers near the junction [ref]. The PL intensity profile of the cross section of the

resultant plates has been used to evaluate the junction width, which is the distance when the PL

intensities of both N=1 and N=2 falls to half of their maximum value (Fig. 2h). The estimated junction width is around 1.1 μm, limited by the spot size of the laser beam and the quench of the emission near the junction due to the carrier separation mentioned above.

The existence of the distinct interface between N=1 and N=2 can be further confirmed by the reflection spectrum of the N=1/N=2 heterostructures (Fig. 2i). Due to the different refractive index of N=1 and N=2, the multiple reflection takes place from the top surface of the N=2 layer and the interface between N=1 and N=2 of the heterostructural single crystals (Supplementary Fig. 3a).

The reflected light beams from the top surface and the interface have a relative fixed phase if the interface is flat enough and has relatively sharp change of refractive index. As a result, the interference takes place, leading to the oscillations in the reflection spectrum below the bandgap regime. Indeed, we observed oscillations on the reflection spectrum at the long wavelength regime, which can be ascribed to the interference due to the top surface and interface. To confirm this, we also measured the reflection spectra for N=1 and N=2 single crystals with similar total thickness (Fig. 2i) and no such oscillation has been observed. This strongly supports that the interface of our heterostructures is rather sharp. Based on the complex-matrix form of the Fresnel equations, we are able to simulate the reflection spectrum of the as-grown heterostructures without taking into account the absorption (Supplementary Fig. 3b). By adding the measured reflection spectra of N=1 and N=2 to the simulated reflection spectrum, we obtained the final simulated reflection spectrum as shown in Fig. 2i, agreeing well with the measured one. The thickness estimated from the oscillation peaks is around 1.46 μm, consistent with the one evaluated from the absorption spectrum according to the Beer–Lambert law (see below). The presence of the oscillations due to the interference reveals that the junction width of our heterostructures is smaller than 70 nm, a quarter of the wavelength in the N=2 perovskites. The resonance of the excited exciton states of N=2 with the trap states of N=1 enables us to observe the first (2s) and second (3s) excited exciton states of N=2.

To further verify our hypothesis, we intentionally tune the mass ratio of BAI to MAI while keeping other reaction conditions the same to examine whether the N>2 2D perovskite can be further synthesized on the resultant heterostructures and how the thickness of N=2 changes with the mass ratio. Surprisingly, XRD pattern and PL spectra indicate that the absence of N>2 2D perovskites and no substantial thickness difference of the N=2 layer have been observed for the resultant heterostructures when the mass ratio of BAI to MAI changes from 1:4 to 1:64 (Fig. 2b and c). For the ratio of 2:3 case, the thickness change might be due to the change of the maintaining time (See below). The similar thickness of the N=2 layer in the resultant heterostructures under different mass ratio of BAI/MAI reveals that the reaction kinetics for the growth of N=2 might be diffusion control. The absence of N>2 2D perovskites for different BAI to MAI ratio proves our hypothesis that the solubility of N>2 2D perovskites is rather high that the aqueous solution has not reached the saturation condition for N>2 2D perovskites to precipitate (Fig. 1c).

In order to further verify that the diffusion process controls the growth of N=2 layer we hypothesized, we have intentionally tuned the time maintaining at 75 ºC when the N=2 2D perovskite starts to grow with a fixed BAI to MAI ratio of 2:3. As shown in Fig. 3a, it is expected that the thickness of N=2 layer (along the out-of-plane direction) would increase with prolonging the maintaining time at 75 ºC since MA cations would diffuse further deep into the n=2 plates with time provided that the process is diffusion control. We have carried out a series of experiments with different maintaining time (at 75 ºC, a constant MAI concentration of 1 Mmol and a fixed BAI to MAI ratio of 2:3) to investigate how the thickness evolves with time through X-ray diffraction, room-temperature absorption spectra and PL spectra (Fig. 3b-d). XRD patterns of the as-grown N=1/N=2 show that only diffraction peaks from N=1 and N=2 2D perovskites are present with a longer maintaining time, suggesting the absence of N>2 perovskites (Fig. 3b). Nevertheless, the intensity of diffraction peaks from N=2 increases with the increase of the maintaining time, which implies that the thickness of N=2 layer increases along with the maintaining time. As the maintaining time increases, the intensity of absorption edge locating at 520 nm (corresponding to absorption edge of N=1) gradually decreases while the intensity of the absorption edge at 590 nm (corresponding to absorption edge of N=2) continuously increases indicating the continuous thickness increase of the N=2 layer with the maintaining time, which finally reaches a constant value according to the Beer–Lambert law (Fig. 3c, e). The maximum thickness of N=2 layer is estimated to be around 1.5 μm, which is probably limited by the finite diffusion length at a certain temperature and a certain concentration of MAI. The evolution of the room-temperature PL spectra with the maintaining time exhibits the same trend as the XRD patterns and absorption spectra. The intensity of emission peak at 520 nm continuously decreases with prolonging the maintaining time at 75 ºC while the intensity of peak at 590 nm continuously increases, indicating the gradual increase of the thickness of the N=2 layer (Fig. 3d). In particular, we have found that the measured thickness of the as-grown heterostructural single crystal plates maintains around 80 μm (Fig. 3e) no matter what the lateral size of the plates is, how we prolong the maintaining time and how we tune the ratio of BAI to MAI, which unambiguously indicates that the growth of N=2 layer is indeed after the formation of the N=1 plates and the growth is a diffusion controlled process, as we hypothesized in Fig. 1b and c. To sum up, XRD diffraction patterns, room-temperature absorption and PL spectra all point to that N=2 layer grows thicker and no additional N>2 perovskite compositions has been formed with increasing the maintaining time, which eventually confirm that the reaction kinetics of N=2 layer is a diffusion-control process.

In addition, the total thickness of the heterostructural single crystals can be readily tuned by intentionally varying the MAI concentration of the reaction solution but with a fixed BAI to MAI ratio of 2:3 (Fig. 3a). As shown in Fig. 3f, the total thickness nearly linearly increases with the MAI concentration of the reaction solution. We further examined how the change of MAI concentration affects the growth process by reducing the MAI concentration to 0.3 Mmol (but a fixed BAI to MAI ratio of 2:3) and varying the maintaining time at 75 °C. Similar trend as the MAI concnetraion of 1 Mmol was observed: the total thickness of the heterostructures maintains around 20 µm while the thickness of the N=2 layer gradually increases with the maintaining time but with a different growth rate compared with that of 1 Mmol. This suggests that the change of the concentration only tune the total thickness and the growth rate of the N=2 layer without fundamentally altering the growth process. Therefore, the diffusion controlled growth process together with tuning the concentration of the reaction solution enables us to conveniently prepare large-size 2D perovskite heterostructural single crystals with controlled thickness and junction depth for further fundamental investigations and exploring potential applications in the field of electronics and optoelectronics.

As a demonstration, electronic devices based on those centimeter-size heterostructural single crystals have been fabricated to investigate their optoelectronic properties. Figure 4a displays the schematic illustration of two-probe vertical electronic devices for the N=1/N=2 heterostructural single crystals. Taking advantage of the centimeter size of the heterostructures, we can directly transfer the as-growth samples grown under different maintaining time to an indium-tin-oxide (ITO) substrate (also acting as bottom electrodes) and define the 10-nm Cr/100-nm Au top electrodes by shadow mask and e-beam evaporation. In this device architecture since the N=1 plate are sanwiched by N=2 layers, the device is actually consisted by two back-to-back heterostructures, which can efficiently suppress the dark current and improve the on-off ratio. Based on previous studies, the band alignment diagram of the our heterostructures can be drawn, which belongs to type-II structure (Fig. 4b)[42]. This band alignment also favors the photogenerated carrier separation and thus the detection efficiency[43].

Figure 4c exhibits the output characteristics of the device in dark and under different monochromatic light illuminations for the heterostructural plates with a total thickness of ~80 μm (the maintaining time is 240 min, corresponding to a thickness of ~1.1 μm for N=2 layer). Due to the high crystalline quality of our single crystal plates (a very low intrinsic carrier concentration) and the back-to-back heterostructure device structure we adopted, an extremely small dark current of ~$10^{-12}$ A was observed, which is on the same order of single crystal N=2 photodetectors[44], and even two orders of magnitude lower than that of graphene/N=1 photodetectors[45]. Under illumination, the notable photocurrent appears and varies with incident light wavelength. The asymmetric current-voltage curve indicates that the contact is not optimal. Unlike the 3D perovskites, there is no observable hysteresis in the output characteristics probably due to the insulating organic layers, which prohibits the ion migration in the out-of-plane direction[46]. Interestingly, the spectral response indicates that there are two spectral response bands with the peaks locating at the around 540 nm and 610 nm, which shows around 20 nm redshifts for both peaks compared with the absorption edge of the heterostructures (Fig. 4d). This implies that the charge collection narrow mechanism (or surface recombination) would be the possible reason for the narrow dual-band spectral response[47,48]. It should be noted that both spectral response peaks are close to excitonic absorption peaks for N=1 and N=2 respectively, which indicates the excitonic absorption contributes to the photocurrent. This is surprising since the exciton binding energy in 2D perovskites is on the order of hundreds of meV and thus we don't expect the excitons can be ionized by the thermal excitation with a thermal energy of ~25 meV at room temperature. Nevertheless, the exciton ionization is further supported by the PL intensity, which is greatly reduced in the heterostructures compared with that of N=1 and N=2 (Supplementary Fig. 4)[27]. One possible reason for the exciton ionization here is the built-in field formed within the heterostructures, but further investigations are required to clarify this.

To exclude the influence from the variation of light intensity at different wavelength, we have adopted responsivity R (defined as the ratio of the photocurrent and the incident light intensity) rather than photocurrent in Fig. 4d. Strikingly, FWHMs for those two response bands are 20 nm for green band (540 nm) and 35 nm for red band (610 nm), both of which are narrower than the commercial fluorescence band-pass filters (40-90 nm)[49]. The corresponding maximum responsivity (external quantum efficiency, denoted as EQE) for those two response bands are 0.69 AW$^{-1}$ (158 %) and 0.65 AW$^{-1}$ (132 %) for green band (540 nm) and red band (610 nm), both of which are almost three orders of magnitude larger than that of 3D perovskite narrow band photodetectors[47]. Furthermore, the detectivity (D*) characterizing the weakest level of light that the device can detect is calculated by D*=$RA^{1/2}/(2qI_{dark})^{1/2}$, where A is the detector area, $I_{dark}$ is the dark current and q is the absolute value of electron charge. The estimated D* value of the device is $6*10^{10}$ Jones, which is comparable with the reported value of 2D perovskite photodetectors[50].

Figure 4e shows the optical switch characteristics of the device at different negative voltages under a 540-nm light illumination. The absolute value of the photocurrent continuously increases with the increase of the absolute value of the applied voltage, consistent with the output characteristics in Fig. 4c. The on and off current maintains almost the same for all measured cycles suggesting the excellent stability and reversibility of our devices. The extracted on-off ratio is around $10^3$ under the applied voltage of -3 V, which is also larger than the reported value[47, 50]. The response speed, characterized by rise time and decay time, is an important parameter for the photodetectors and optical switches. The rise time and decay time of the photodetector are defined as the time taken for the photocurrent increasing from 10 % to 90 % of the peak value, and vice versa. The measured rise time and decay time of our devices are 150 ms and 170 ms illuminated by the 540-nm light under the applied voltage of 3 V (Fig. 4f), which are slower than the reported 3D perovskite based narrow band photodetectors. This slow response speed might be due to the thick plates (80 μm) and smaller electric field we applied (0.03 V/μm).

Furthermore, we have fabricated the photodetectors based on the heterostructures with a thinner N=2 layer of around 300 nm by controlling the maintaining time at 75 ºC to be 30 minutes in order to examine how the thickness of the N=2 layer affects the performance of the narrow dual-band photodetections. Interestingly, the output characteristic for the thinner N=2 layer exhibits a notable rectifying behavior with forward-to-reverse bias current ratios of $2 \times 10^3$ and both the dark current and photocurrent dramatically increases compared with that of the device with the thicker N=2 layer mentioned above (Supplementary Fig. 5a). Under illumination with different monochromatic light, significant photocurrent on the order of 100 nA has been observed under a forward bias of 3 V. The optical switch characteristic also indicates the excellent stability and reversibility similar to the device with the thicker N=2 layer but with much larger dark current (Supplementary Fig. 5b). We have measured several devices and all exhibit the similar behavior. The diode behavior and enhanced dark current and photocurrent might originate from the depletion field of the thinner N=2 layer. The low carrier concentration of our single crystals would lead to a large depletion length, which might exceed the thickness of the thinner N=2 layer and thus affect the contact barrier. Due to the different contact materials for the bottom electrodes (ITO) and top electrodes (Au), the depletion field might have different degree of effect on the contact barrier, which gives rise to the diode behavior and enhanced current. For the devices with thicker N=2 layer, the N=2 layer has not been completely depleted and thus no such effects are observed. Nevertheless, more investigations are required to clarify the mechanism, which is out of scope of this manuscript.

Importantly, narrow dual-band spectral response with central peaks locating at 540 nm and 610 nm has been observed for the devices with thinner N=2 layer as well with FWHMs of 10 nm at green band and 38 nm at red band (Fig. 4g). The estimated responsivity (EQE) for those two bands are 11.5 AW$^{-1}$ (2640 %) and 22.7 AW$^{-1}$ (4614 %), which are much larger than the devices with thicker N=2 layer. Interestingly, compared with the thicker N=2 layer device, the photoresponse peak related to the absorption band of the N=2 layer shows a slight blueshift while the responsivity at the valley between the 540 nm and 610 nm are much higher suggesting that the thickness of N=2 layer is too thin (150 nm) to prevent the photogenerated carrier from diffusing away from surface, similar to the 3D perovskite narrow band photodetectors[32,47]. Nevertheless, there is no such change for the photoresponse peak associated with N=1 layer for the thin N=2 layer device since in both devices the N=1 layer are sufficiently thick. This thickness dependence of spectral response further strengthens that charge collection narrowing mechanism (or surface recombination) would be very likely to be the primary reason for us to observe the narrow dual-band spectral response in our heterostructure devices[47].

To demonstrate that our synthetic strategy and narrow dual-band photodetectors is general for a wide range of 2D perovskite heterostructures other than $(BA)_2(MA)_{N-1}Pb_NI_{3N+1}$, we have successfully synthesized the centimeter-size $(BA)_2PbBr_4/(BA)_2(MA)Pb_2Br_7$ and $(PEA)_2PbI_4/(PEA)_2(MA)Pb_2I_7$ 2D perovskite heterostructural single crystals by applying the synthetic method we developed and fabricated photodetectors based on them. XRD patterns and room-temperature PL spectra confirm the formation of the $(BA)_2PbBr_4/(BA)_2(MA)Pb_2Br_7$ (Fig. 5a and c) and $(PEA)_2PbI_4/(PEA)_2(MA)Pb_2I_7$ (Fig. 5b and d). For comparison, we also display the XRD pattern and PL spectra of $(BA)_2PbBr_4$ and $(BA)_2(MA)Pb_2Br_7$ in Fig. 5a and c, and $(PEA)_2PbI_4$ and $(PEA)_2(MA)Pb_2I_7$ in Fig. 5c and d, all of which are synthesized by solution method[39]. The output characteristics of the photodetectors based on $(BA)_2PbBr_4/(BA)_2(MA)Pb_2Br_7$ and $(PEA)_2PbI_4/(PEA)_2(MA)Pb_2I_7$ heterostructural single crystals (Supplementary Fig. 6) show the same trend as that of $(BA)_2PbI_4/(BA)_2(MA)Pb_2I_7$ devices. For the $(BA)_2PbBr_4/(BA)_2(MA)Pb_2Br_7$ devices, due to the thinner $(BA)_2(MA)Pb_2Br_7$ layer (supporting by the very weak emission peak of $(BA)_2(MA)Pb_2Br_7$ compared with that of $(BA)_2PbBr_4$ in Fig. 5c), the output curves both in dark and under monochromatic light illumination show a large rectifying behavior while the linear output curves have been observed for $(PEA)_2PbI_4/(PEA)_2(MA)Pb_2I_7$ devices because of the thicker $(PEA)_2(MA)Pb_2I_7$ layer (supporting by the strong emission of $(PEA)_2(MA)Pb_2I_7$ compared with that of $(PEA)_2PbI_4$ in Fig. 5d). The excellent stability and reversibility of those devices have been demonstrated by the optical switch characteristics in Fig. 5e. Strikingly, for both types of devices the narrow dual-band spectral response is present, ranging from blue to red wavelength range with all FWHMs smaller than 30 nm and fairly good EQE of >10 % (Fig. 5f). By properly selecting the chemical compositions of 2D perovskite heterostructures, we expect the narrow dual-band photodetectors can be tuned continuously from the ultraviolent range to near infrared range.

The layered characteristic of the 2D perovskites in nature allows the heterostructures to be seamlessly grown along the out-of-plane direction while the lattice mismatch between N=1 and N=2 2D perovskites within the in-plane direction prohibits the growth of the mixed N=1 and N=2 perovskites inside the structures which grants the formation of N=1 perovskite is sanwiched by N=2 heterostructural single crystal plates. This diffusion-control growth process also ensures the purity and high crystalline quality of the resultant heterostructures, resulting in the excellent performance of the highly narrow dual-band photodetectors in the entire visible wavelength range. The excellent stability of 2D perovskites against the moisture, oxygen and light illumination has inherently equipped within our synthesized heterostructures. After being stored under ambient condition for 150 days, XRD patterns and PL spectra retain their original diffraction peaks without any impurity peaks, a strong indicator of the excellent stability of our heterostructures (Supplementary Fig. 7). The excellent performance together with stability of our heterostructures makes our narrow dual-band photodetectors more attractive for multicolor imaging sensing in a variety of applications.

In summary, we have developed a novel solution method to controllably synthesize centimeter-size 2D perovskite heterostructural single crystals with tunable thickness and junction depth and realized highly narrow dual-band photodetections with excellent performance based on those as-grown heterostructural single crystals. The layered nature of 2D perovskites and diffusion-control growth process allow us to achieve seamless growth of the heterostructural single crystals with high crystalline quality and high phase purity. The high crystalline quality and high pure phase of our as-synthesized heterostructural single crystals enable us to achieve narrow dual-band photodetection by the charge collection narrow mechanism with excellent performance. Our studies not only open up an avenue to controllably synthesize high quality, pure phase 2D perovskite based heterostructures for fundamental investigations and potential electronic and optoelectronic applications, but also provide an alternative approach to realize optical-filter-free ultraviolet, visible or infrared narrow dual-band photodetection for a variety of applications related to multicolor imaging technology.

## Methods

**Synthesis of MAI/BAI/BABr/PEAI**. The precursor MAI was synthesized by adding methylamine (40 % wt.% in $H_2O$) into HI (57 wt.% in $H_2O$) aqueous solution with a molar ratio of 1:1 drop by drop. The resultant solution was stirred for 2 hrs at a constant temperature of 0 °C. The solution was then heated to 60 °C to evaporate the solvent and the remaining solid was washed by cold diethyl ether three times followed by drying at 70 °C for 12 hrs. For synthesis of BAI/BABr/PEAI solution, the same procedure was used except that the methylamine was replaced by n-butylamine (phenethylamine) and HI by HBr and the stirring time was extended to 4 hrs.

**Synthesis of centimeter-size single crystal 2D perovskite heterostructural plates.** 0.5 g PbO powder was dissolved in a mixture of 3 ml HI (57 % w/w aqueous) and 0.5 ml $H_3PO_2$ (50 % w/w aqueous) by heating to 140 °C under constant magnetic stirring. Then 1.75 mmol BAI solution was added into the resultant solution and slowly cooled down to 75 °C while 0.2 g MAI powder was successively injected into the solution for the synthesis of $(BA)_2(MA)PbI_4/(BA)_2(MA)Pb_2I_7$ plates. Afterwards, the stirring was stopped, and the solution was maintained at 75 °C for different time according to the demanding, and naturally cooled down to room temperature within about 30 minutes.

For the synthesis of $(BA)_2PbBr_4/(BA)_2(MA)Pb_2Br_7$ ($(PEA)_2PbI_4/(PEA)_2(MA)Pb_2I_7$) plates, the exact same procedures were adopted except replacing the 57 % w/w aqueous HI solution (BAI solution) with 48 % w/w aqueous HBr solution (PEAI solution).

**Material characterizations.** Powder X-ray diffraction measurements were recorded using Bruker D2 PHPSER (Cu Kα λ=0.15419 nm, Nickel filter, 25 kV, 40 mA). Optical microscopy images were collected by the Olympus BX53M system microscope. The scanning electron microscopy (SEM) images were acquired on a JEOL 7001F field emission scanning electron microscope. The thickness of the heterostructures was measured on a stylus profiler (Dektak XT, Bruker). The absorption spectra were recorded on an ultraviolet-Vis spectrophotometer (UV-1750, SHIMADZU). The photoluminescence measurements were performed in backscattering configuration using a Horiba HR550 system equipped with a 600 g/mm grating excited by a 473 nm solid-state laser with a power of 1 μW. The reflection measurement was carried out in a Horiba HR550 system with a 600 g/mm grating illuminated by a 100 W halogen tungsten lamp.

**Photoconductivity measurements.** 10-nm Cr/100-nm Au electrodes were defined by a shadow mask and deposited by electron beam evaporation. A quartz tungsten halogen lamp (250 W) was used as the light source and then dispersed by a monochromator (Horiba JY HR320). The monochromatic light beam output was used to illuminate the devices mounted on a chip carrier. The incident light intensity was recorded by a pyroelectric detector (Gentec, model APM (D)). The photocurrent was collected by a lock-in amplifier (Stanford SR830) coupled with a mechanical chopper while the response speed was acquired on a digital oscilloscope (Tektronix MDO3032). All measurements were performed at ambient atmosphere.

# Reference

1.      Egger, D. A. *et al.* What remains unexplained about the properties of halide perovskites? *Adv. Mater*. 1800691 (2018).

2.      García de Arquer, F. P., Armin, A., Meredith, P. & Sargent, E. H. Solution-processed semiconductors for next-generation photodetectors. *Nat. Rev. Mater.* **2,** 16100 (2017).

3.      Huang, J., Yuan, Y., Shao, Y. & Yan, Y. Understanding the physical properties of hybrid perovskites for photovoltaic applications. *Nat. Rev. Mater.* **2,** 17042 (2017).

4.      Ha, S.-T., Shen, C., Zhang, J. & Xiong, Q. Laser cooling of organic–inorganic lead halide perovskites. *Nat.Photon.* **10,** 115-121 (2015).

5.      Huang, H. *et al.* Colloidal lead halide perovskite nanocrystals: synthesis, optical properties and applications. *NPG Asia Mater.* **8,** 328 (2016).

6.      Jeon, N. J. *et al.* Solvent engineering for high-performance inorganic-organic hybrid perovskite solar cells. *Nat. Mater.* **13,** 897-903 (2014).

7.      Xing, G. *et al.* Low-temperature solution-processed wavelength-tunable perovskites for lasing. *Nat. Mater.* **13,** 476-480 (2014).

8.      Huang, J., Tan, S., Lund, P. D. & Zhou, H. Impact of $H_2O$ on organic–inorganic hybrid perovskite solar cells. *Energy Environ. Sci.* **10,** 2284-2311 (2017).

9.      Li, X. *et al.* Improved performance and stability of perovskite solar cells by crystal crosslinking with alkylphosphonic acid omega-ammonium chlorides. *Nat. Chem.* **7,** 703-711 (2015).

10.     Koh, T. M., Febriansyah, B. & Mathews, N. Ruddlesden-Popper perovskite solar cells. *Chem* **2,** 326-327 (2017).

11.     Liao, Y. *et al.* Highly oriented low-dimensional tin halide perovskites with enhanced stability and photovoltaic performance. *J. Am. Chem. Soc.* **139,** 6693-6699 (2017).

12.     Etgar, L. The merit of perovskite's dimensionality; can this replace the 3D halide perovskite? *Energy Environ.Sci.* **11,** 234-242 (2018).

13.     Shi, E. *et al.* Two-dimensional halide perovskite nanomaterials and heterostructures. *Chem. Soc. Rev.* DOI: 10.1039/C7CS00886D (2018).

14.     Quan, L. N. *et al.* Ligand-stabilized reduced-dimensionality perovskites. *J. Am. Chem. Soc.* **138,** 2649-2655 (2016).

15.     Zhang, Q., Chu, L., Zhou, F., Ji, W. & Eda, G. Excitonic properties of chemically synthesized 2D organic-inorganic hybrid perovskite nanosheets. *Adv. Mater*. 1704055 (2018).

16.     Blancon, J. C. *et al.* Extremely efficient internal exciton dissociation through edge states in layered 2D perovskites. *Science* eaal4211 (2017).

17.     Quan, L. N. *et al.* Tailoring the energy landscape in quasi-2D halide perovskites enables efficient green-light emission. *Nano Lett.* **17,** 3701-3709 (2017).

18.     Booker, E. P. *et al.* Formation of long-lived color centers for broadband visible light emission in low-dimensional layered perovskites. *J. Am. Chem. Soc.* **139,** 18632-18639 (2017).

19.     Ma, L., Dai, J. & Zeng, X. C. Two-dimensional single-layer organic-inorganic hybrid perovskite semiconductors. *Adv. Energy Mater.* **7,** 1601731 (2017).

20.     Grancini, G. *et al.* One-Year stable perovskite solar cells by 2D/3D interface engineering. *Nat. Commun.* **8,** 15684 (2017).

21.     Dou, L. T. *et al.* Atomically thin two-dimensional organic-inorganic hybrid perovskites. *Science* **349,** 1518-1521 (2015).

22.     Lotsch, B.V. Vertical 2D Heterostructures. *Annu. Revi. Mater. Res.* **45,** 85-109 (2015).

23. Duan, X., Wang, C., Pan, A., Yu, R. & Duan, X. Two-dimensional transition metal dichalcogenides as atomically thin semiconductors: opportunities and challenges. *Chem. Soc. Rev.* **44,** 8859-8876 (2015).

24. Liu, Y. *et al.* Van der Waals heterostructures and devices. *Nat. Rev. Mater.* **1,** 419-425 (2016).

25. Novoselov, K. S., Mishchenko, A., Carvalho, A. & Castro Neto, A. H. 2D materials and van der Waals heterostructures. *Science* **353,** aac9439 (2016).

26. Sulpizio, J. A., Ilani, S., Irvin, P. & Levy, J. Nanoscale phenomena in oxide heterostructures. *Annu. Rev. Mater. Res.* **44,** 117-149 (2014).

27. Wang, J. *et al.* Controllable synthesis of two-dimensional Ruddlesden-Popper-type perovskite heterostructures. *J. Phys. Chem. Lett.* **8,** 6211-6219 (2017).

28. Wang, Y. *et al.* Visible light driven type II heterostructures and their enhanced photocatalysis properties: a review. *Nanoscale* **5,** 8326-8339 (2013).

29. Zhou, N. *et al.* Narrowband spectrally selective near-infrared photodetector based on up-conversion nanoparticles used in a 2D hybrid device. *J. Mater. Chem. C* **5,** 1591-1595 (2017).

30. Sang, L., Hu, J., Zou, R., Koide, Y. & Liao, M. Arbitrary multicolor photodetection by hetero-integrated semiconductor nanostructures. *Sci. Rep.* **3,** 2368 (2013).

31. Liu, H. C. *et al.* GaAs/AlGaAs quantum-well photodetector for visible and middle infrared dual-band detection. *Appl. Phys. Lett.* **77,** 2437-2439 (2000).

32. Wang, P. *et al.* Arrayed van der Waals broadband detectors for dual-band detection. *Adv. Mater.* **29,** 1604439 (2017).

33. Liu, X. *et al.* All-printable band-edge modulated ZnO nanowire photodetectors with ultra-high detectivity. *Nat. Commun.* **5,** 4007 (2014).

34. Pan, L., Ishikawa, A. & Tamai, N. Detection of optical trapping of CdTe quantum dots by two-photon-induced luminescence. *Phys. Rev. B* **75,** 161305 (2007).

35. Rao, H. S., Li, W. G., Chen, B. X., Kuang, D. B. & Su, C. Y. In situ growth of 120 cm$^2$ $CH_3NH_3PbBr_3$ Perovskite crystal film on FTO glass for narrowband-photodetectors. *Adv. Mater.* **29,** 1602639 (2017).

36. Sobhani, A. *et al.* Narrowband photodetection in the near-infrared with a plasmon-induced hot electron device. *Nat. Commun.* **4,** 1643 (2013).

37. Geyer, S. M., Scherer, J. M., Moloto, N., Jaworski, F. B. & Bawendi, M. G. Efficient luminescent down-shifting detectors based on colloidal quantum dots for dual-band detection applications. *ACS Nano* **5,** 5566-5571 (2011).

38. Stoumpos, C. C. *et al.* High members of the 2D Ruddlesden-Popper halide perovskites: synthesis, optical Properties, and solar cells of $(CH_3(CH_2)_3NH_3)_2(CH_3NH_3)_4Pb_5I_{16}$. *Chem* **2,** 427-440 (2017).

39. Stoumpos, C. C. *et al.* Ruddlesden–Popper hybrid lead iodide perovskite 2D homologous semiconductors. *Chem. Mater.* **28,** 2852-2867 (2016).

40. Mitzi, D. B., Dimitrakopoulos, C. D. & Kosbar, L. L. Structurally tailored organic-inorganic perovskites: optical properties and solution-processed channel materials for thin-film transistors. *Chem. Mater.* **13,** 3728-3740 (2001).

41. Liang, K., Mitzi, D. B. & Prikas, M. T. Synthesis and characterization of organic-inorganic perovskite thin films prepared using a versatile two-step dipping technique. *Chem. Mater.* **10,** 403-411 (1998).

42. Cao, D. H., Stoumpos, C. C., Farha, O. K., Hupp, J. T. & Kanatzidis, M. G. 2D homologous perovskites as light-absorbing materials for solar cell applications. *J. Am. Chem. Soc.* **137,** 7843-7850 (2015).

43. Peng, B. *et al.* Ultrafast charge transfer in $MoS_2/WSe_2$ p–n Heterojunction. *2D Materials* **3,** 025020 (2016).

44. Li, L. *et al.* Tailored engineering of an unusual $(C_4H_9NH_3)_2(CH_3NH_3)_2Pb_3Br_{10}$ two-dimensional multilayered perovskite ferroelectric for a high-performance photodetector. *Angew. Chem. Int. Ed.* **56,** 12150-12154 (2017).

45. Tan, Z. *et al.* Two-dimensional $(C_4H_9NH_3)_2PbBr_4$ perovskite crystals for high-performance photodetector. *J. Am. Chem. Soc.* **138,** 16612-16615 (2016).

46. Yuan, Y. & Huang, J. Ion migration in organometal trihalide perovskite and its impact on photovoltaic efficiency and stability. *Acc. Chem. Res.* **49,** 286-293 (2016).

47. Fang, Y., Dong, Q., Shao, Y., Yuan, Y. & Huang, J. Highly narrowband perovskite single-crystal photodetectors enabled by surface-charge recombination. *Nat. Photon.* **9,** 679-686 (2015).

48. Lin, Q., Armin, A., Burn, P. L. & Meredith, P. Filterless narrowband visible photodetectors. *Nat. Photon.* **9,** 687-694 (2015).

49. Park, H. *et al.* Filter-free image sensor pixels comprising silicon nanowires with selective color absorption. *Nano Lett.* **14,** 1804-1809 (2014).

50. Chen, S. & Shi, G. Two-dimensional materials for halide perovskite-based optoelectronic devices. *Adv. Mater.* **29,** 1605448 (2017).

51. Li, D. H. *et al.* Size-dependent phase transition in methylammonium lead iodide perovskite microplate crystals. *Nat. Commun.* **7,** 11330 (2106).

52. Wang, G. M. *et al.* Wafer-scale growth of large arrays of perovskite microplate crystals for functional electronics and optoelectronics. *Sci. Adv.* **1,** 1500613 (2015).

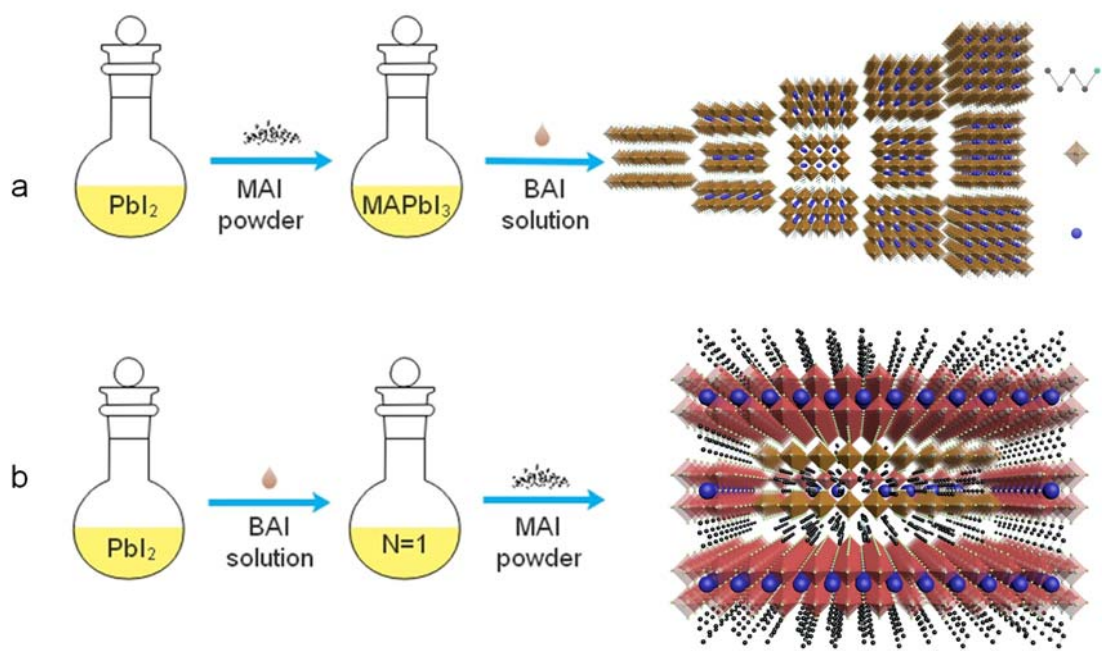

**Figure 1 | Synthetic strategy of 2D perovskite heterostructures**. (**a**) Schematic of solution synthetic process of $(BA)_2(MA)_{N-1}Pb_NI_{3N+1}$ (N=1-5) 2D perovskites. (**b**) Schematic illustrations of synthesis of $(BA)_2PbI_4/(BA)_2(MA)Pb_2I_7$ 2D perovskite heterostructural single crystals.

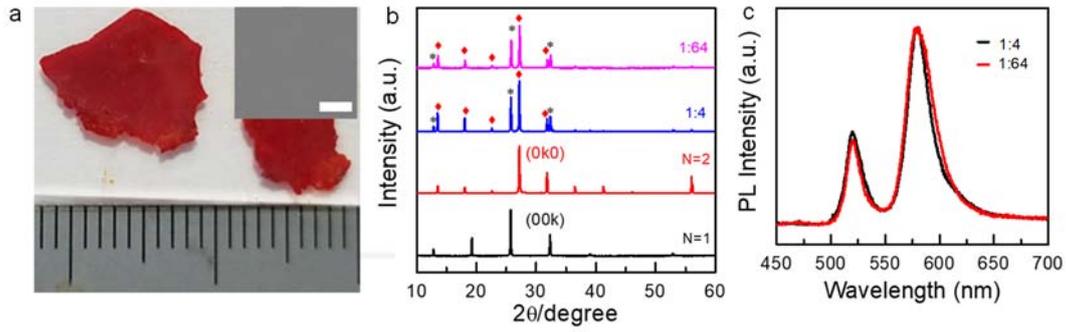

**Figure 2 | Structural characterization of (BA)$_2$PbI$_4$/(BA)$_2$(MA)Pb$_2$I$_7$ heterostructural single crystals.** (**a**) Optical image of (BA)$_2$PbI$_4$/(BA)$_2$(MA)Pb$_2$I$_7$ heterostructures grown with a maintaining time of 240 min. Inset: SEM image of the as-grown heterostructural single crystals. The scale bar is 20 μm. (**b**) XRD patterns of N=1, N=2 obtained by solution method and as-grown (BA)$_2$PbI$_4$/(BA)$_2$(MA)Pb$_2$I$_7$ heterostructural single crystals with different mass ratio of BAI/ MAI. The asterisk peaks belong to (BA)$_2$PbI$_4$ while rhombus peaks can be indexed to (BA)$_2$MAPb$_2$I$_4$. (**c**) Normalized PL spectra of 2D perovskite heterostructural single crystals with different mass ratio of BAI/ MAI.

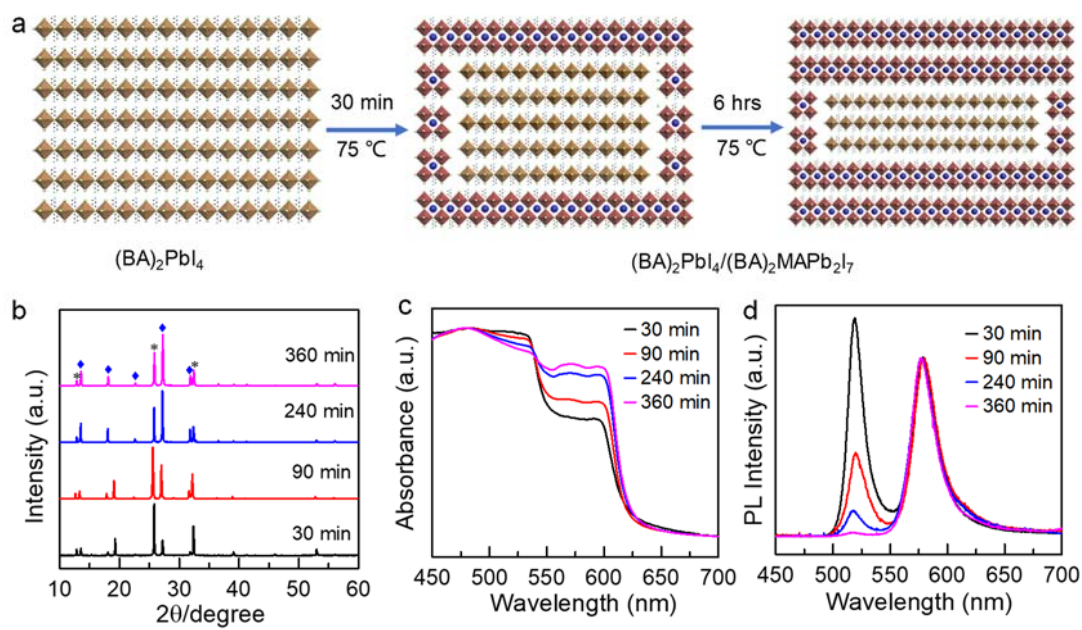

**Figure 3 | Growth mechanism of centimeter-size (BA)₂PbI₄/(BA)₂(MA)Pb₂I₇ heterostructural single crystals.** (**a**) Schematic illustrations of the growth process of the $(BA)_2(MA)Pb_2I_7$ layer outside the $(BA)_2PbI_4$ plates with the increase of the maintaining time at 75 °C. (**b**) XRD patterns of as-grown $(BA)_2PbI_4/(BA)_2(MA)Pb_2I_7$ heterostructural single crystals with different maintaining times. The asterisk peaks belong to $(BA)_2PbI_4$ while rhombus peaks can be indexed to $(BA)_2MAPb_2I_4$. (**c**, **d**) Normalized absorption spectra (c) and PL spectra (d) of $(BA)_2PbI_4/(BA)_2(MA)Pb_2I_7$ heterostructures with different maintaining times. The thickness of $(BA)_2(MA)Pb_2I_7$ layer can be evaluated from the absorption spectra (Figure S2).

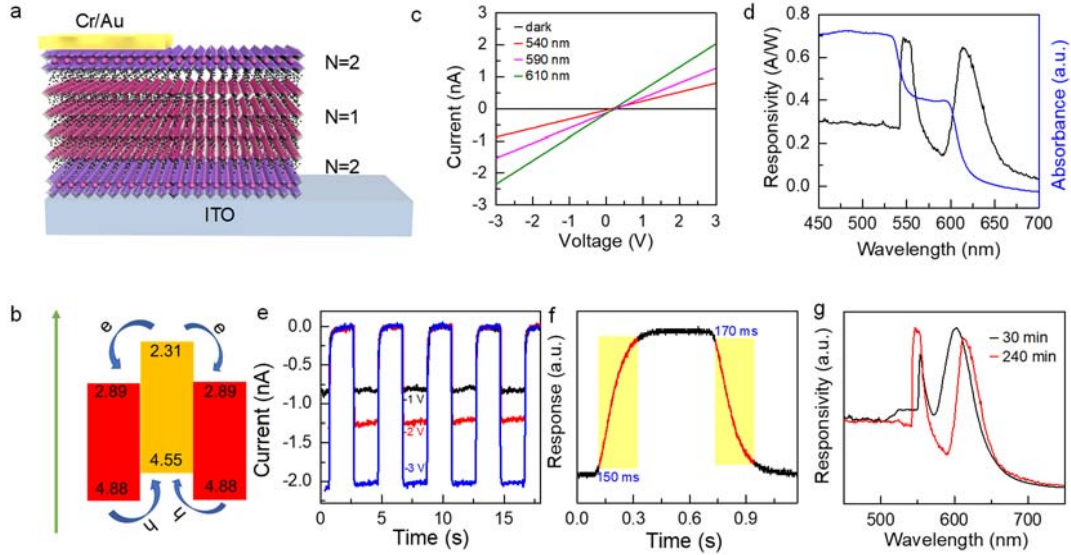

**Figure 4 | Highly narrow dual-band photodetectors.** (**a**) Schematic of device configuration of the $(BA)_2PbI_4/(BA)_2(MA)Pb_2I_7$ heterostructural single crystals. (**b**) Band diagram of the corresponding $(BA)_2PbI_4/(BA)_2(MA)Pb_2I_7$ heterostructural single crystal based on the previously reported UPS data. (**c**) Output characteristics (*I* versus *V*) of the heterostructure devices in dark and under different monochromatic light illumination. (**d**) The corresponding responsivity and normalized absorbance versus incident wavelength of the $(BA)_2PbI_4/(BA)_2(MA)Pb_2I_7$ heterostructural single crystal photodetector under the bias of 3 V. (**e**) Optical switch characteristics of the $(BA)_2PbI_4/(BA)_2(MA)Pb_2I_7$ heterostructural single crystal photodetector under different bias voltage from -1 to -3 V illuminated by a 540 nm light with a power density of 45 mW/cm$^2$. (**f**) Temporal photocurrent response of the $(BA)_2PbI_4/(BA)_2(MA)Pb_2I_7$ heterostructural single crystal photodetector illuminated by a 540 nm light with a power density of 45 mW/cm$^2$ under a bias voltage of 3 V. (**g**) Normalized responsivity for the narrow dual-band photodetectors of $(BA)_2PbI_4/(BA)_2(MA)Pb_2I_7$ heterostructures with the maintaining time of 30 min and 240 min (corresponding to a $(BA)_2(MA)Pb_2I_7$ layer thickness of 300 nm and 1.1 μm, respectively) under a bias voltage of 3 V.

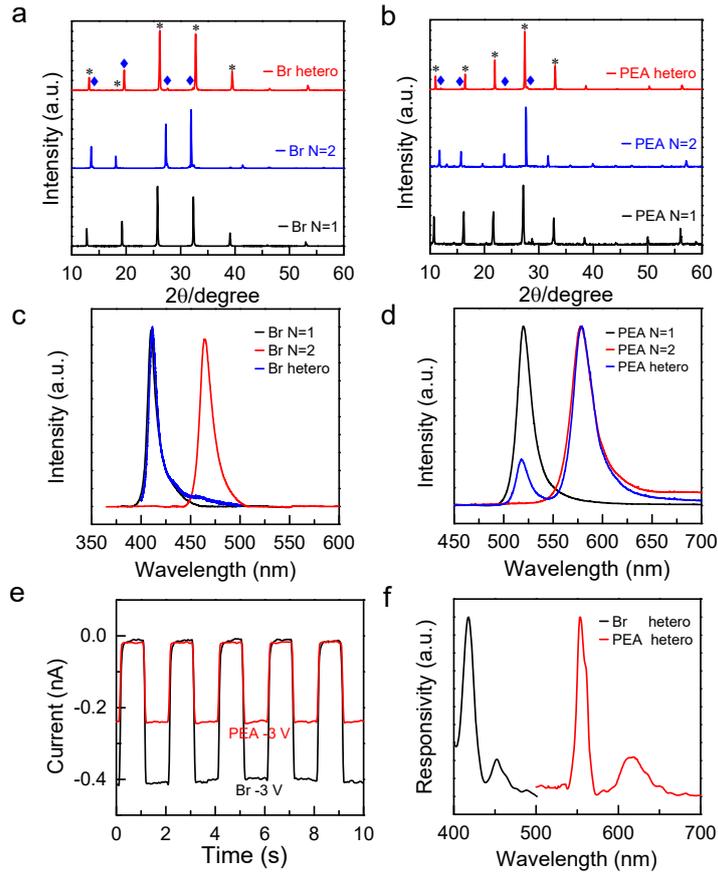

**Figure 5 | Narrow dual-band photodetectors based on (BA)₂PbBr₄/(BA)₂MAPb₂Br₇ and (PEA)₂PbI₄/(PEA)₂MAPb₂I₇ heterostructural single crystals.** (**a, b**) XRD patterns of (BA)₂PbBr₄/(BA)₂MAPb₂Br₇ (a) and (PEA)₂PbI₄/(PEA)₂MAPb₂I₇ (b) heterostructures. For comparison, XRD patterns of their corresponding N=1 and N=2 2D perovskites synthesized by solution method are shown in corresponding figures as well. (**c, d**) PL spectra of (BA)₂PbBr₄/(BA)₂MAPb₂Br₇ (c) and (PEA)₂PbI₄/(PEA)₂MAPb₂I₇ (d) heterostructures. For comparison, PL spectra of their corresponding N=1 and N=2 2D perovskites synthesized by solution method are displayed as well. (**e**) Optical switch characteristics of the (BA)₂PbBr₄/(BA)₂MAPb₂Br₇ device under a 410-nm light (24 mW/cm²) illumination and of the (PEA)₂PbI₄/(PEA)₂MAPb₂I₇ device under a 540-nm (46 mW/cm²) illumination at a bias voltage of -3 V. (**f**) Spectral response of the (BA)₂PbBr₄/(BA)₂MAPb₂Br₇ and (PEA)₂PbI₄/(PEA)₂MAPb₂I₇ based highly narrow dual-band photodetectors under a bias voltage of 3 V with the spectral response in the entire visible wavelength range.